\documentclass[conference]{IEEEtran}

\newcommand{\Eurora}{Eurora} 

\newcommand{\Cineca}{CINECA} 

\newcommand{\revised}[1]{{\color{black}#1}}

\usepackage{graphicx}
\usepackage{color}

\begin{document}

\title{Predicting System-level Power for a Hybrid Supercomputer}

\author{\IEEEauthorblockN{Alina S\^irbu}
\IEEEauthorblockA{Department of Computer Science\\
University of Pisa, Italy\\
Email: alina.sirbu@unipi.it}
\and
\IEEEauthorblockN{Ozalp Babaoglu}
\IEEEauthorblockA{Department of Computer Science and Engineering\\
University of Bologna, Italy\\
Email: ozalp.babaoglu@unibo.it}
}

\maketitle

\begin{abstract}
For current High Performance Computing systems to scale towards the holy grail of ExaFLOP performance, their power consumption has to be reduced by at least one order of magnitude. This goal can be achieved only through a combination of hardware and software advances. Being able to model and accurately predict the power consumption of large computational systems is necessary for software-level innovations such as proactive and power-aware scheduling, resource allocation and fault tolerance techniques. In this paper we present a 2-layer model of power consumption for a hybrid supercomputer (which held the top spot of the Green500 list on July 2013) that combines CPU, GPU and MIC technologies to achieve higher energy efficiency. Our model takes as input workload information --- the number and location of resources that are used by each job at a certain time --- and calculates the resulting system-level power consumption.
When jobs are submitted to the system, the workload configuration can be foreseen based on the scheduler policies, and our model can then be applied to predict the ensuing system-level power consumption. Additionally, alternative workload configurations can be evaluated from a power perspective and more efficient ones can be selected.
Applications of the model include not only power-aware scheduling but also prediction of anomalous behavior.
\end{abstract}
~
\\

%


\section{Introduction}
Power consumption of computational systems has become a major concern worldwide. Today, it is not uncommon for a large data center to consume as much power as a mid-size city, with the obvious economic and environmental consequences.  Furthermore, large power needs have negative implications for the systems themselves, for example by limiting their scalability. Accurate modeling power consumption for large computing systems will be extremely important for optimizing their energy usage. Models allow for prediction of system behavior in various scenarios, enabling advanced scheduling and fault tolerance techniques that are essential for making Exascale computing sustainable. 

In this paper we model and predict system-level power consumption starting from workload measures for Eurora~\cite{cavazzoni2012}, an experimental hybrid High Performance Computing (HPC) installation packing CPUs, GPUs and MICs that was ranked \#1 on the Green500 list in July 2013. Prediction is obtained in two steps. First, we develop a relation between power used by \emph{computing components} and the \emph{total system} power, including networking, IO system and other components. Second, power consumption of computing components is predicted from workload data based an earlier study where we introduced a Support Vector Regression approach~\cite{sirbu2016}. The two steps are then combined to obtain prediction of system-level power starting from workload measures.

This work brings several contributions to modeling power consumption for HPC systems. First, the relation between computing components and system power is investigated, and a clear linear dependency between the two is observed, \revised{in agreement with other studies. We take this result one step further by building a complete power model for the entire system. Our second contribution is thus a 2-layer model} that is capable of predicting system-level power starting from workload data. The method does not require knowledge of application code or hardware counters for power prediction.  \revised{This makes the model easily extendable to other systems, since only simple workload measures that are common to all HPC systems are used.} Third, possible applications of the model are investigated including power-aware scheduling and failure prediction based on changes in model fit.

The rest of the paper is organized as follows.  State-of-the-art is surveyed in Section~\ref{sec_related} followed by a discussion of the data and our prediction approach in Sections~\ref{sec_data} and \ref{sec_prediction}. Results of the first modeling step --- from component to system power --- are presented in Section~\ref{sec_sys1}. The second step --- from workload to component power consumption --- is investigated in Section~\ref {sec_comp} and the two steps are integrated in Section~\ref{sec_sys2}. Section~\ref{sec_disc} discusses potential applications and Section~\ref{sec_concl} summarizes and concludes the paper.

\section{Related work}\label{sec_related}
With energy needs becoming a major concern for large computational infrastructures, numerous recent research efforts have focused on reducing power usage. A large amount of work regards modeling power for various types of computing units, starting from load, frequency and other hardware counters. For instance, single and dual core CPU power is modeled in~\cite{Dargie2015} by considering the relation between the probability distribution functions of load and power, while servers with up to 8 cores are studied in~\cite{Takouna2011,Kim2014}. GPU power is estimated from load measures in~\cite{Ma2009}.  These methods do not allow for advance prediction in real life scenarios, since load and hardware counters cannot be known in advance, unless they can be predicted through other methods. Our method is significantly different in that we model total system power starting exclusively from workload measures, without the need to monitor the individual components, enabling \emph{advance prediction of power}.

Power of HPC applications has also been analyzed in recent years. For instance, the US Department of Defense are using application signatures to predict power consumption across different architectures~\cite{Olschanowsky2010}. Performance counters are used to model application power on three small scale HPC platforms by~\cite{Witkowski2013}. GPU CUDA kernels are analyzed in~\cite{Nagasaka2010}, again based on job performance counters. Recently, we have introduced a method~\cite{sirbu2016} based on Support Vector Regression (SVR), which builds one power model per user, to predict job power consumption based on workload in Eurora. This method has an advantage over others in that it does not require instrumenting the applications to extract signatures and performance counters, but only needs the number of resources required, making it much more straightforward to apply. In this work, this SVR method will be employed to predict power of computing components, from which we will then obtain system-level power.

In the quest for Exascale computing systems, 
where energy needs will be much greater,
it is important to analyze power of large computing infrastructures \emph{at system level}. Related work cited above looks only at individual computing components or jobs, while we concentrate on total system power including hardware other than computing components, such as networking and I/O. Very few other examples of power analysis at system level exist in the literature, despite a recognized need for development in this direction~\cite{dayarathna2015}. For example, recently Google has introduced a method~\cite{Gao2014} of modeling Power Usage Effectiveness (PUE) through an Artificial Neural Network which takes as input workload, cooling, power, together with other external information such as outside temperature, wind speed, etc. This allowed for testing various data center scenarios and improving PUE for the system under analysis.

Predictions of power consumption at system level are useful to enforce power-aware policies on the computational systems. Power capping is one way of managing increasing power needs of computational infrastructures. For example, the authors in~\cite{Borghesi2015, Borghesi2015a} introduce a scheduling procedure that takes power into consideration. However, they look only at the power of computing components and estimate power usage only on average for each component type. Studies along these lines could benefit from accurate system-level power prediction that we introduce in this paper.

A different application of power models is prediction or identification of anomalous behavior. For instance, the Google PUE model~\cite{Gao2014} allowed for identification of anomalies in monitoring, when the model did not fit the data any more. Similarly, a decrease in modeling performance can predict system failures as well. Our model can be used also to predict anomalous behavior, as we will discuss in section~\ref{sec_disc}.

\section{{\Eurora} data}\label{sec_data}

{\Eurora} is a hybrid HPC system installed at \Cineca, the largest data center in Italy~\cite{cineca}. It was built as a prototype for energy efficiency research and topped the Green500 list in July 2013. The machine consists of one rack of 64 nodes, combining three types of computing components: CPUs (8-core Intel Xeon E5 CPUs), GPUs (Nvidia Tesla Kepler) and MICs (Intel Xeon Phi). Each node is equipped with two CPUs and either two GPUs or two MICs.  All nodes run CentOS Linux. The workload is handled through the Portable Batch System (PBS), while cooling is based on hot liquid technology. A custom monitoring framework~\cite{bartolini2014} stores detailed logs related to the load and power consumption of each individual component, together with power and workload at system level in a MySql database. The database contains measurements from March 2014 to August 2015, with over 250GB of data in 328 tables.

\begin{figure}[!b]
\centering
\includegraphics[width=0.45\textwidth]{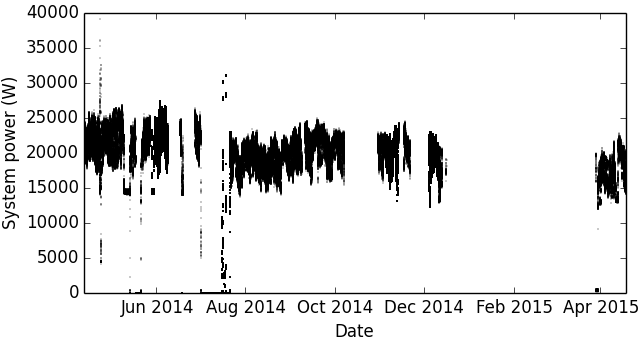}
\caption{Availability of power measurements at system level.}
\label{fig_pow}
\end{figure}

\begin{figure*}[!t]
\centering
\includegraphics[width=0.65\textwidth]{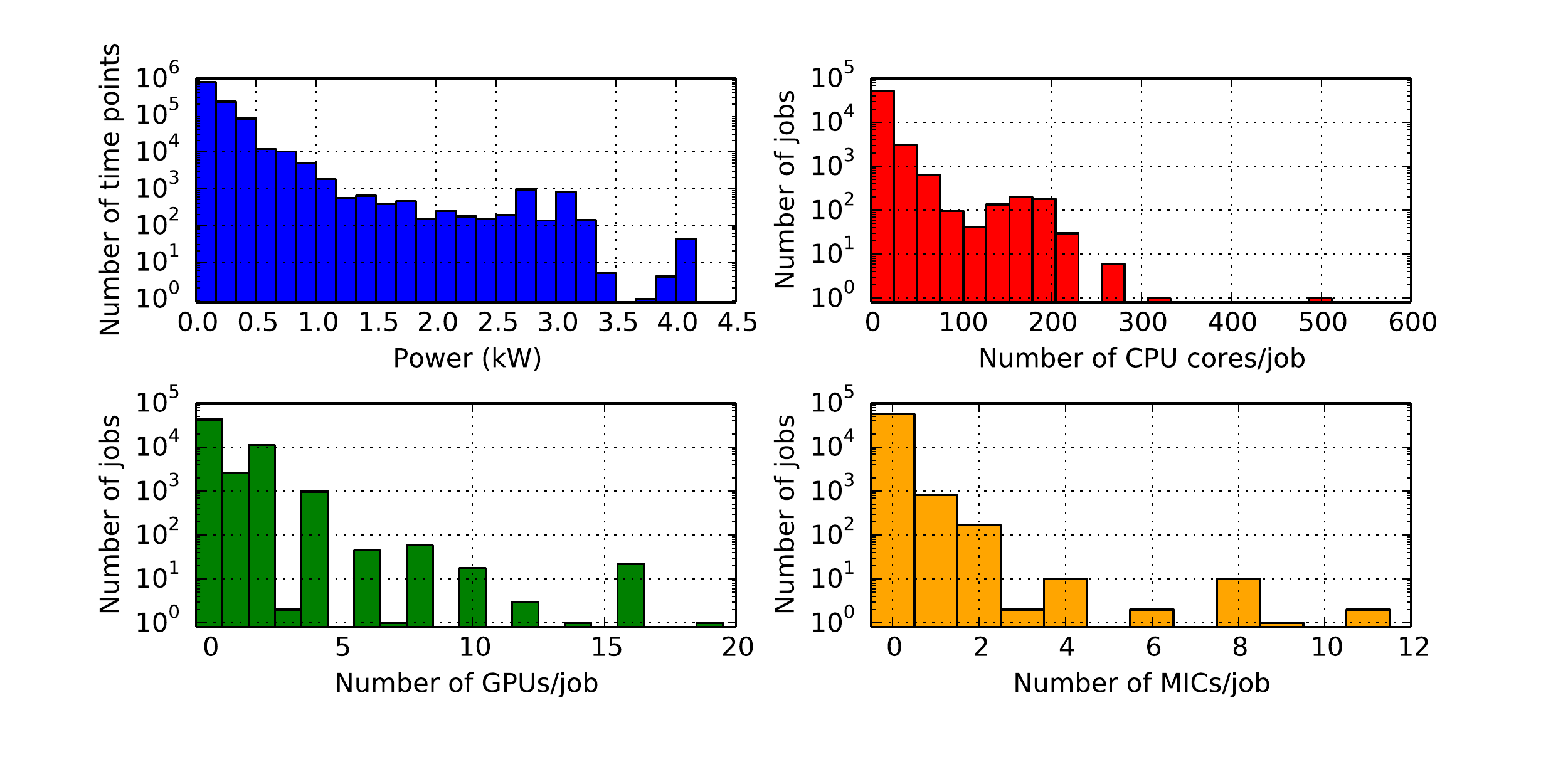}
\caption{\revised{Workload properties for corrected data. The top-left plot shows the distribution of power consumption per job, measured at 5 minute intervals. The rest of the histograms show the distribution of resources used by each job (CPU cores, GPUs and MICs). }}
\label{fig_workload}
\end{figure*}

Workload information is necessary to understand what jobs are running in the system and the number of resources they are allocated at each node. The database contains a table including all the jobs (405,756 unique jobs), with information about the user (401 unique users), job name, start and end timestamp. A different table matches each job to all the nodes it uses, including the exact number of CPU cores, GPUs and MICs used on these nodes. By combining these two tables, we were able to compute how many resources were used by each job on each node at a 5-minute resolution.

Another crucial piece of information for the study was the power consumption of computing components. For each node, the database contains three tables describing the statuses of the CPUs, GPUs and MICs present at the node. This includes power consumption of the components collected at 5-second resolution generating large amounts of data. On average, CPU tables contain over 11M rows, GPU tables over 6M rows and MIC tables over 800K rows. This difference in the table sizes is due to the fact that some components are shut down more often than others (for instance MICs are shut down when unused, while CPUs only when they fail). It is important to note that power data is known only at the level of CPU, GPU and MIC, and is not available at the core level.
For each node, we consider the power at 5-minute intervals for the two CPUs, GPUs, and MICs, i.e., a subset of the 5-second power data. We consider power at time $t$ to be the measurement that was performed between $t-4s$ and $t+5s$ that is closest in time to $t$ (e.g., to compute power at 01:00:00, data in the interval 00:59:56 -- 01:00:05 is considered and  the measurement closest in time is selected). \revised{These measurements were used to compute total power of computing components, but also total power for each job.}

\begin{figure*}[!t]
\centering
\includegraphics[width=0.63\textwidth]{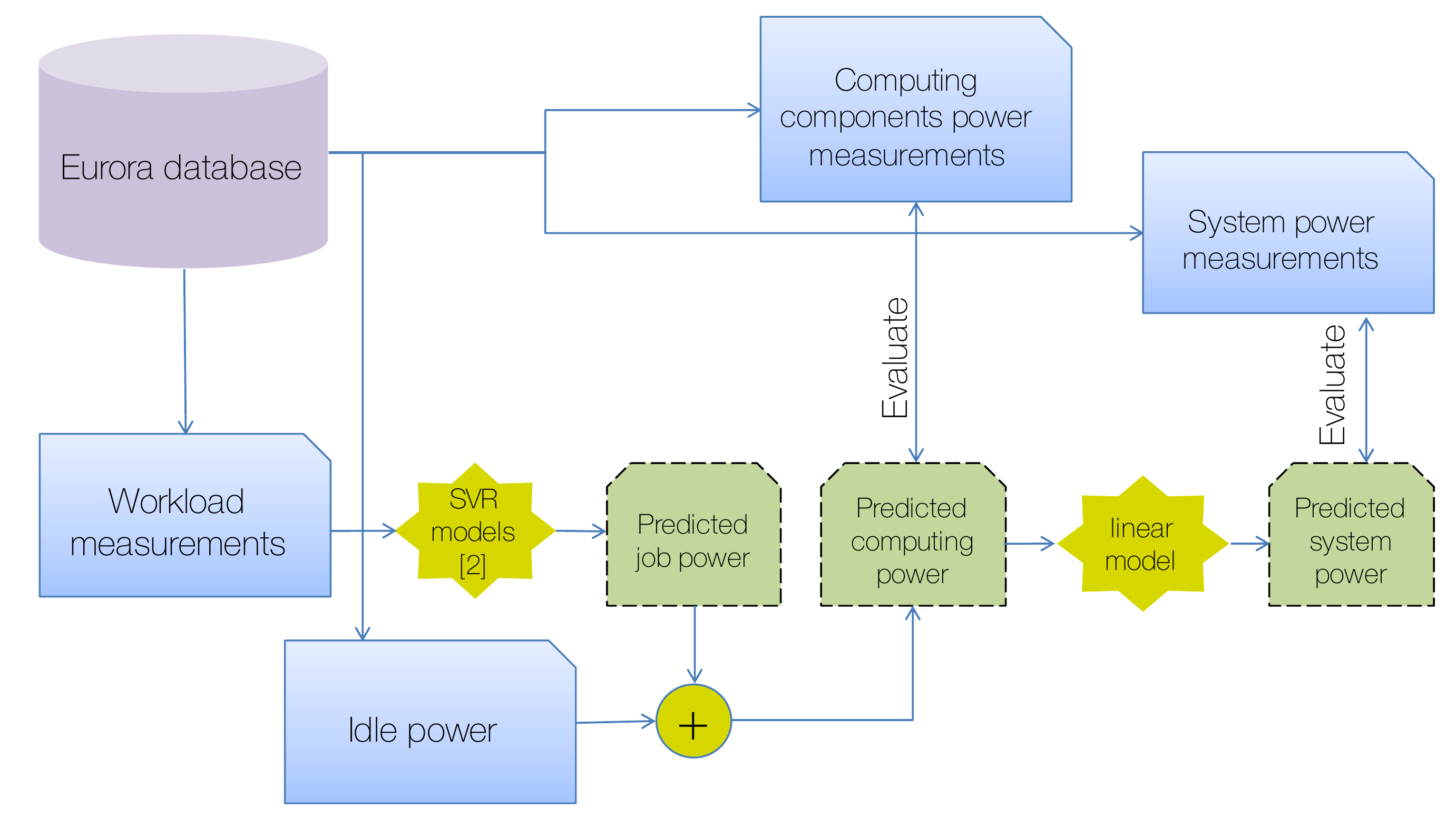}
\caption{Our layered modeling approach.}
\label{fig_model}
\end{figure*}

A third piece of information extracted from the database is total system-level power usage, measured outside the {\Eurora} system through the main electric panel powering the installation. This accounts for the power consumed by the entire rack and not only the CPU, GPU and MIC as discussed above, and is the final target of our prediction study. Figure~\ref{fig_pow} shows power measurements available at this level. Several gaps appear, due to both system shutdown and monitoring issues. For this study we concentrate on system-level data during the period from July 2014 to November 2014, when {\Eurora} was more stable and we could obtain enough contiguous data for training and testing our models. 

Before extracting the features of interest, we studied the data to identify inconsistencies. The main issue identified was missing data points for various types of measurements (CPU, GPU, MIC power or job information), reducing drastically the overlapping period for the various data types available. Given that each type of measurement is performed by a different daemon running on each node, we devised techniques to correct them, in order to obtain a dataset large enough for our analysis. For this purpose, if no power measurement was available for a particular node (neither CPU, GPU, MIC, nor jobs), we considered the node to be down and its power to be null. However, if any of the power measurements was available, it meant the node was up. In this case, if no jobs seemed to use the components on this node, then the missing power measurement was set to the idle value. However, if jobs were running on the specific component, the power could not be estimated, so the data point was eliminated altogether. After all corrections were performed, component-level data was synchronized with system level power measurements. This resulted in a total of 99,825 valid time points sampled at 5 minute intervals \revised{for system level power, with 57,183 jobs for the 401 users in the system.} The corrected data was the basis of the prediction approach presented in the next section. \revised{Figure~\ref{fig_workload} shows the distribution of power levels recorded at 5 minute intervals for each job, together with
 the distribution of 
 the number of components used by each job. Many jobs use only CPUs with about 26\% of jobs employing GPUs and about 2\% employing MICs}.

\section{Prediction approach}\label{sec_prediction}

Using the data described above, we predict power consumption at system level based only on workload measures. This involves two steps as we will outline below. Figure~\ref{fig_model} presents graphically the approach adopted.

\begin{figure}[!b]
\centering
\includegraphics[width=0.4\textwidth]{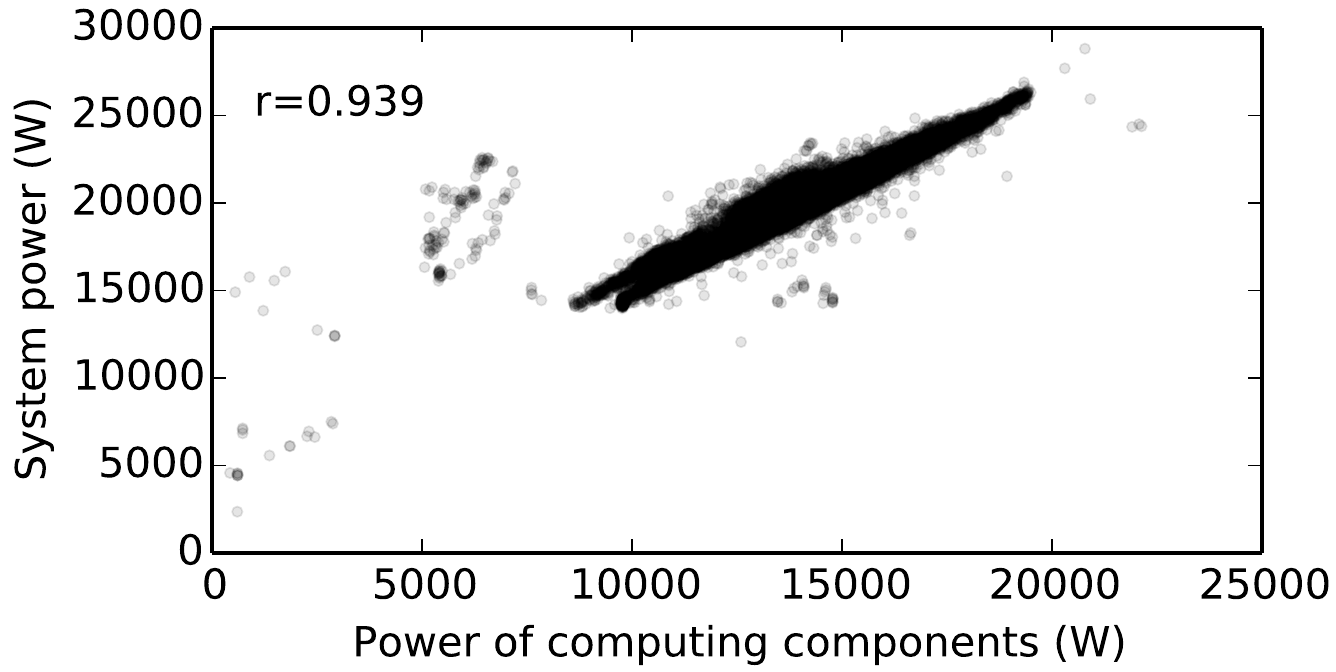}
\caption{Power consumed by the entire system versus power of computing components only.}
\label{fig_data}
\end{figure}

\begin{figure*}[!ht]
\centering
\includegraphics[width=0.65\textwidth]{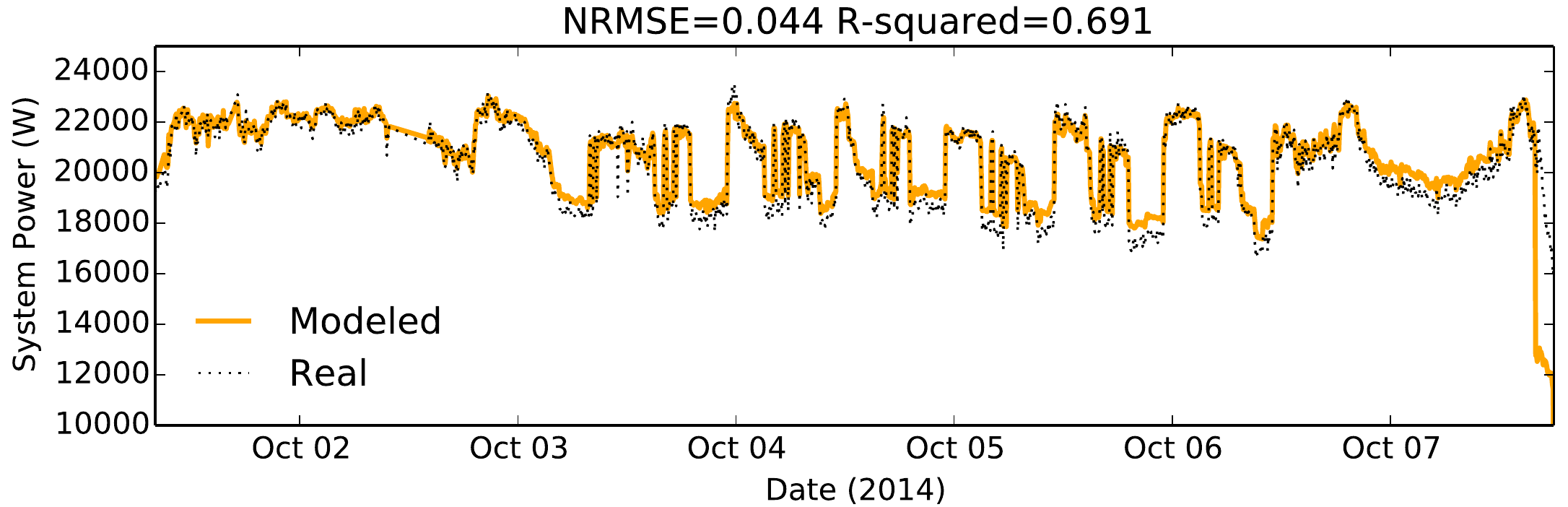}
\caption{Power consumed by the entire system, estimated, using a linear model, from power of computing components. The dashed line shows the real (measured) system-level power consumption. }
\label{fig_sys1}
\end{figure*}

The first step of our analysis is to build a model of \emph{system-level power consumption}, starting from the power of the computational components. We extracted measurements of power consumption at system level at 5-minute intervals from the {\Eurora} database. Power of each individual CPU, GPU and MIC was also extracted and summed to obtain total computing power. Figure~\ref{fig_data} plots power of the system versus total computing power for our dataset. It is clear that there is a strong relationship between the two power measures, with a very high Pearson correlation coefficient (0.939). Thus, we built a \emph{linear model} of system power, denoted as $LM$, starting from the individual computing components. The model provides an \emph{estimate} of the system level power $P_S^*(t)$ at time $t$ as a linear function of the \emph{measured} power of computing components $P_C(t)$ at the same time $t$:

\begin{equation}\label{eq_s1}
P_S^*(t)=LM(P_C(t))
\end{equation}

In order to evaluate the model, we opted for a classical cross-validation approach, where the model is trained and tested on separate datasets. In the following we will show results from training with all data from September 2014 and testing on data from October 2014.

The second step of our analysis is predicting \emph{power consumption of computing components} starting from workload measures. This enables power prediction in advance for a certain workload. The analysis was based on our previous work that allows predicting \emph{power consumption of jobs} in {\Eurora}, using only job characteristics and collocation information~\cite{sirbu2016}. The method uses Support Vector Regression (SVR) to build one model per user. For each job, a wide set of regression features are employed to predict its power profile in time. A job is described by independent features, which are job name, number of CPU cores, GPUs and MICs used by the job and number of nodes allocated, but also by features that describe the workload and resource allocation globally, for example the number of cores/GPUs/MICs in use by other jobs collocated on the same nodes as the job being analyzed. This choice enables prediction of power interference across jobs, so that different mapping of resources can result in differences in power levels. The analysis is based on workload and power data extracted from the {\Eurora} database. Specifically, we use measurements of component power to compute the exact power used by each job at 5 minute intervals (the regression target) while features are extracted from the information about resource allocation that exists in the database. Further details of job power prediction with this approach can be found in the original paper~\cite{sirbu2016}.

Having obtained job power profiles (predicted at 5-minute intervals using our SVR method), we compute the total \emph{predicted} power of computing components $P_C^*(t)$ at time $t$  by summing the \emph{predicted} power of individual jobs running on the system at time $t$, $P_j^*(t)$, together with the power of the idle components, $P_{idle}(t)$:

\begin{equation}
P_C^*(t)=\sum_{j \in \mathrm{Jobs}}{P_j^*(t)}+P_{\mathrm{idle}}(t)
\end{equation}

Idle power can be measured once for each component type, with $P_\mathrm{idle}$ being the sum over all idle components. In the following we will show results for October 2014, where SVR models were trained with data before October and applied to compute total power for October 2014. 

We only applied the SVR method to users with at least 1000 data points for training, coming from at least 100 different jobs. For users who had less data available, we used an \emph{Enhanced Average Model} (EAM), also introduced in~\cite{sirbu2016}. For this, we computed, for each user $u$, an average power per component type (CPU core: $\bar{P}_{CPU}^u$, GPU: $\bar{P}_{GPU}^u$, MIC: $\bar{P}_{MIC}^u$) from the limited amount of existing training data. For each job $j$ belonging to the user, we considered the number of components used by each job: $n_{CPU}^j$, $n_{GPU}^j$ and $n_{MIC}^j$. The predicted job power can then be computed as:
\begin{equation}
P_j^*=n_{CPU}^j \times \bar{P}_{CPU}^u + n_{GPU}^j \times \bar{P}_{GPU}^u+ n_{MIC}^j \times \bar{P}_{MIC}^u
\end{equation}
This value was used at all time points $t$ when the job was active. In the rare case where no user data for training existed, we used a global (over all users) average power consumption per component in the EAM.

The two models of steps one and two can be then combined in a \emph{2-layer model}, to obtain the desired \emph{system-level power predictions}:

\begin{equation}
P_S^*(t)=LM(P_C^*(t))=LM\left(\sum_{j \in \mathrm{Jobs}}{P_j^*(t)}+P_{\mathrm{idle}}(t)\right)
\end{equation}
It is important to note that the linear model $LM$ is trained using the \emph{real} (measured) power consumption of components, $P_C(t)$ in Equation~\ref{eq_s1}, but in the final model it is applied to the \emph{predicted} power of components, $P_C^*(t)$. 
Again, application of the model on test data from October 2014 will be shown below. Both linear regression and SVR were performed using the \emph{scikit-learn} Python package~\cite{scikit-learn}, while data preprocessing for feature extraction was performed using the BigQuery cloud platform~\cite{bigquery}. All analysis scripts are available on GitHub~\cite{github}.

At each step, the models were evaluated using two standard criteria for regression:  the (mean-)normalized-root-mean-squared-error (NRMSE) and  R-squared ($R^2$). 

\begin{equation}
\makebox{NRMSE}=\frac{\sqrt{ (\sum_{i=1}^N{(P_S(t_i)-P_S^*(t_i))^2}) / {N}}} {\bar{P}_S}
\end{equation}

\begin{equation}
R^2=1-\frac{\sum_{i=1}^N(P_S(t_i)-P_S^*(t_i))^2}{\sum_{i=1}^N(P_S(t_i)-\bar{P}_S)^2}
\end{equation}
where $N$ is the number of time points considered, $P_S^*(t_i)$ and $P_S(t_i)$ are the predicted
and real system-level powers at time $t_i$, respectively,
while $\bar{P}_S$ is the average of the real system power over all $N$ data points.

NRMSE measures the error between  prediction and real data as a fraction of the average measured power. It takes positive values only, with small values meaning errors are much smaller than the average power levels. The $R^2$ criterion includes information on variability in the data and compares the errors to the natural variability. It tells us how the model performs compared to the so-called ``average model'' --- a model where power is predicted to be the average of all power levels measured. A value close to 1 means the model performs very well, while a value close to 0 indicates the model is no better than the average model (error is comparable to the standard deviation of the data).


\section{Results}
In this section we discuss the modeling performance at each analysis step introduced in the previous section.

\subsection{From computing components to system-level power}\label{sec_sys1}

\begin{figure*}[!ht]
\centering
\includegraphics[width=0.65\textwidth]{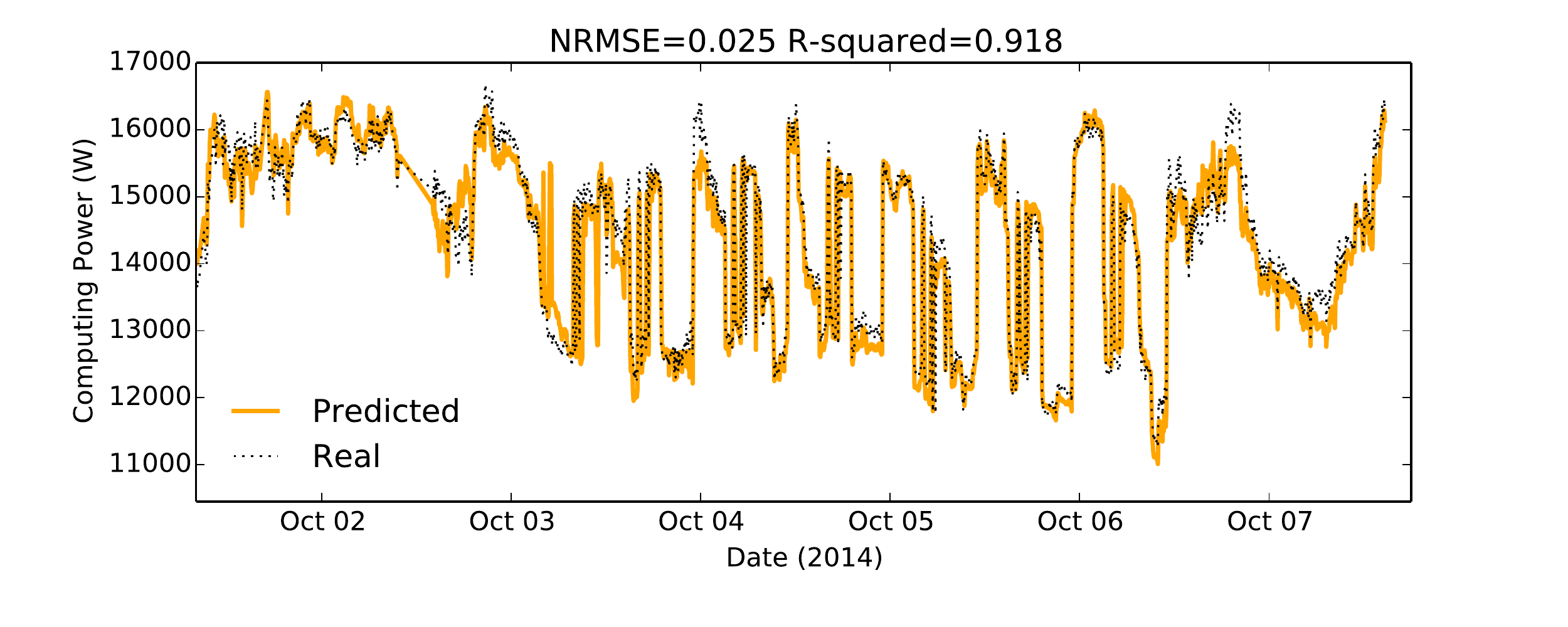}
\caption{Power consumed by computing components, predicted from workload measures. The dashed line shows the real (measured) power of computing components.}
\label{fig_comp}
\end{figure*}

\begin{figure*}[!ht]
\centering
\includegraphics[width=0.65\textwidth]{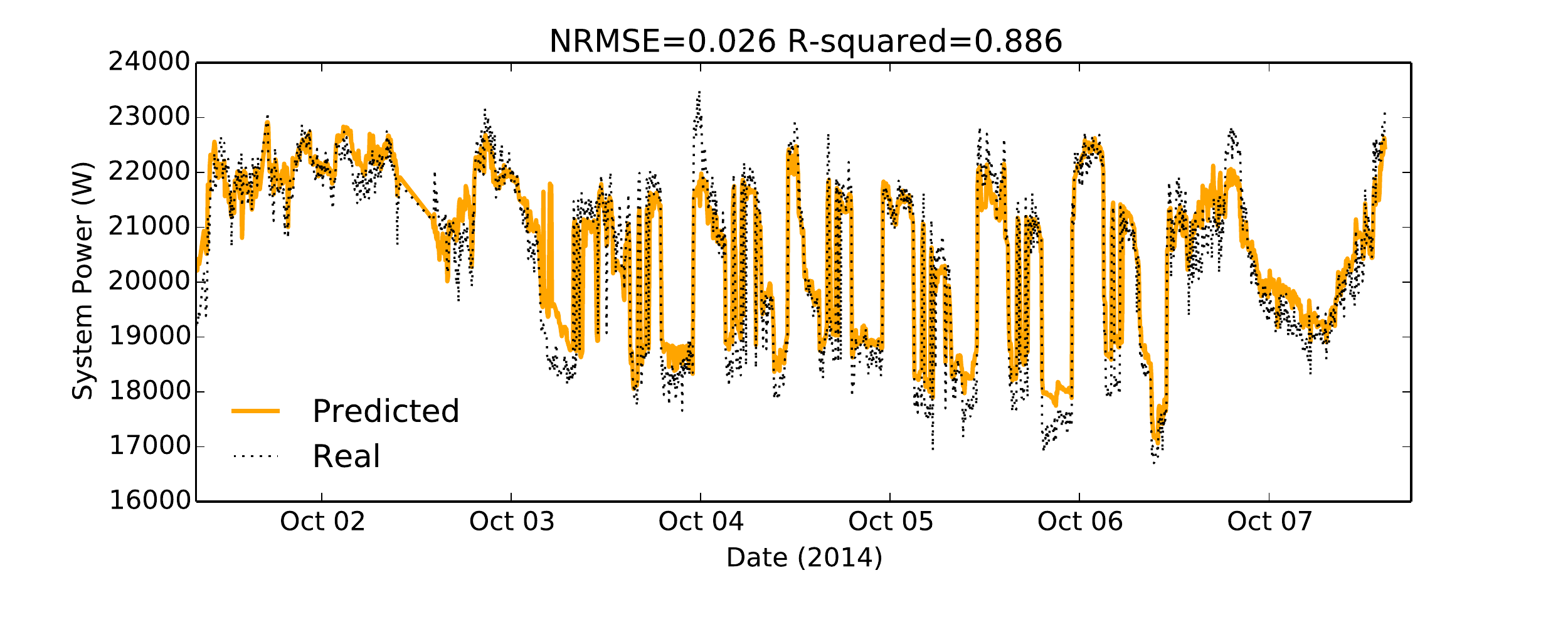}
\caption{Power consumed by the entire system, forecasted with the linear model from predicted computing power extracted from workload measures. The dashed line shows the real (measured) system-level power consumption.}
\label{fig_sys2}
\end{figure*}

The first analysis step was performed on {\Eurora} data from September and October 2014. In particular, a linear model was built to describe power consumption of the entire {\Eurora} machine based on the power measurements for computing components. For cross-validation, the model was trained with data from September 2014 and tested on data from October 2014.

Figure~\ref{fig_sys1} shows the real and estimated power time series, with NRMSE and $R^2$ values included. The linear model provides a very good fit, with errors bellow 5\% and high $R^2$ value. This is a strong indication that a linear model can extrapolate very well from component power to system power.

\subsection{From workload to power of computing components }\label{sec_comp}

During the second step of our analysis, we predict power consumption of computing components based on workload measures, by summing job-level predictions. As we have shown in~\cite{sirbu2016}, job-level prediction achieves very good performance for over 80\% of the users in October. Figure~\ref{fig_comp} displays the total predicted power after summing over all users, compared to the measured time series.

The model performance is again extremely good, with errors under 3\% and very high $R^2$ values. Thus, it seems that although prediction is not perfect for some users, considering all of them together allows for errors for some users to be compensated by other users so that final performance is very good. This means that prediction of total power of computing components can be successfully achieved starting from workload measures.

\subsection{From workload to system-level power}\label{sec_sys2}

The final step of our analysis is to combine the two models to obtain system level power predictions from workload measures. We thus apply the linear model evaluated in Section~\ref{sec_sys1} to the predictions shown in Section~\ref{sec_comp}. Figure~\ref{fig_sys2} displays the final prediction result. It is important to note that this prediction does not use any measurement of power at any level, nor any performance measures, but only information on workload (number of components used by jobs for the various users).

As the figure shows, prediction is very similar to the measured system power, with overall errors for October 2014 of under 3\% and very high $R^2$. We can observe a slight trend of underestimating large power levels, especially for singular peaks, and overestimation of very low power levels. The former appears to be due to underestimation at step 2 (Figure~\ref{fig_comp}) while the latter seems to be caused by the linear model overestimating lowest values (Figure~\ref{fig_sys1}).

It is interesting to compare the performance at step 1, when the linear model was applied directly on the \emph{measurements} of the power of computing components, with the final results, where the model is applied on \emph{prediction} of the power of computing components. For this, we need to consider only the common time points.
In Figure~\ref{fig_sys1} the last point in the system power time series is on Oct 7 at 18:00 hours, while in Figure \ref{fig_sys2}, also showing system-level data, the plot stops at time 14:40 on the same day. 
This is due to missing measurements in the job data between 14:40 and 18:00, which caused those time points to be removed from the job power prediction step (they are also missing from Figure~\ref{fig_comp}). If we consider only the common part, we obtain at step 1 NRMSE$=0.02$ and $R^2=0.933$. Hence, by replacing real with predicted computing component power, performance is only slightly decreased, with errors increasing only by 0.6\%. This validates the layered model that we are proposing.

\section{Discussion}\label{sec_disc}

The analysis presented was performed off-line for historical data. In practice, it is intended for on-line use, for instance when using the live monitoring system to extract feature values and computing power predictions in real time for future time windows. As we have discussed in previous work~\cite{sirbu2016}, prediction of job power profiles does not imply a large overhead to the system. Training and applying the linear model has even lower overhead, with negligible running times required, hence all in all, the model is straightforward to employ.

\begin{figure*}[!t]
\centering
\includegraphics[width=0.65\textwidth]{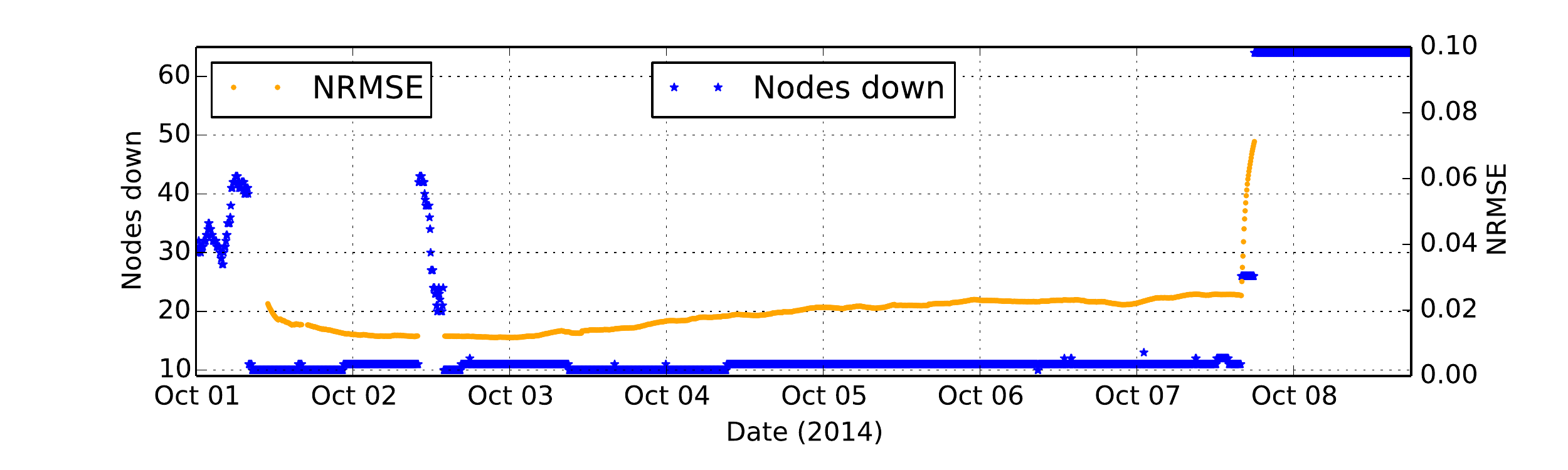}
\caption{Number of down nodes and error of the linear model on a moving 48-hour window over our data. }
\label{fig_error}
\end{figure*}

The future time window for which prediction can be achieved depends on factors such as the frequency of job submission by the user, job length and load of the system. Every time a new job is ready to start, our model can predict the power profile for the system, given also the other jobs currently running on the machine. By predicting job length as well as power (task which we will undertake in the future), a complete system-level profile can be obtained at least until a new job is started. Then the profile changes according to the new job. On a very busy system with long queues, prediction can be obtained for longer periods of time, since we can know in advance what jobs will be scheduled next. On a lightly used system, the prediction window depends practically only on the time between job submission by users, which is more difficult to foresee. HPC systems in general work with relatively long and heavy jobs and queues that are always busy, making our approach very useful in this context.

%
%
\revised{The method presented here is easily applicable to any HPC system. Of course, this involves training models of job power consumption for the users of the new system, and learning the dependencies between power at computing components and system levels, using the methodology outlined in this paper. Once the two model layers are learned, they can be combined into one prediction framework. Thus, on the new system, the required measurements are workload (number of components of each type used by the jobs) and power consumption at component and system level. These are typically available on HPC systems (e.g. \cite{sirbu2015}). When power at component level is missing, temperatures can be used as their proxies~\cite{sirbu2015}. }

System-level power predictions can have several applications for optimization of behavior for HPC systems. Besides providing a tool for operators, to be aware of future power values, the model can be also used to decrease power consumption, by adjusting job scheduling and resource allocation. Power for various scheduling and allocation schemes for the same workload can be predicted, and the scheme with the lowest power employed. This is made possible due to the wide range of workload features considered, which include global description of resources allocated to jobs. Specifically, the number of nodes that a job uses and the number of cores in use by other jobs change from one allocation scheme to another, changing thus power consumption. Our task would be to construct a low-power mapping of resources to jobs, which would require some form of search-based optimization. For instance, methods relying on constraint programming are already widely used for HPC scheduling and could be extended to take into account predictions~\cite{bonfietti2015}. Evolutionary techniques could also be a possibility. The disadvantage of these methods, when facing currently used HPC schedulers, is their running time, hence suitable implementations are needed. 

Another example of a power-aware technique that is commonly explored, especially given the increased power needs of HPC infrastructures, is power capping. This technique is not concerned with decreasing overall energy consumption, although this may happen as a side effect, but concentrates on maintaining total system power at bay, so that the  capacity of the energy provisioning system is not exceeded.  Here too, our model can provide important information on power for future system states, so that scheduling meets the power capping needs.

A different application for our model is prediction of anomalous behavior in the {\Eurora} system. The 2-layer model is trained with data from the monitoring framework, describing a \emph{correctly functioning HPC system}. As described earlier, data was carefully checked for inconsistencies with corrections made when possible and incorrect measurements removed. Hence, the model captures the relation between workload, component power and system power for regular system behavior. If, at some point in time, the system starts to behave differently, the difference between the model output and the real system power will increase. This \emph{different behavior} can appear for several reasons. First, some functioning parameters of the system may have been changed. In this case, models need to be retrained with newest data so that they are up to date. New training sessions can be triggered by administrators shortly after changes to the system have been made. A more interesting case is when this different behavior is caused by anomalies, either in the monitoring system (e.g., wrong or missing measurements, such as the Google data center example~\cite{Gao2014}) or in the HPC system itself.

%
%
%

In order to verify if and how soon the model error increases before a failure, one would require a large number of failure events, possibly annotated with causes,  distinguishing between system and monitoring anomalies. \revised{Even if our dataset is not rich enough} to perform a thorough analysis,  we can analyze the events we could observe. After performing the corrections detailed earlier, we still have measurement gaps that correspond to either the monitoring framework or the system being down. One such period starts on October 7th lasting until October 30th, which indicates a serious problem occurred in the system. We studied the model fit before this system shutdown. For the purposes of this discussion, we provide the fit of the linear model applied to the power of computing components (step 1), since the job data is missing towards the end of the period. However, conclusions can be extended to the complete model. 

Figure~\ref{fig_error} shows the NRMSE of the linear model on a moving window of 48 hours on test data, in October 2015. We also show the number of nodes that appear to be down in the system, based again on missing data. The figure shows clearly how on the 7th of October the number of offline nodes goes to 64, so the entire system is offline. The model error shows a very steep increase about two hours in advance. This sharp steady growth can be considered an advance alarm signaling that the system is starting to fail. Another gap in the data exists on the 2nd of October, however only part of the nodes went down and the system recovered. The model fit does not change before this event, which may signify that the gap could correspond to shutdown of the monitoring system in some nodes, rather than the nodes themselves. 
\revised{We also studied the months of August and September, with only one similar long term system shutdown (about 2 days) preceded by a rise in NRMSE with a 5-minute lead time. Importantly, no such sharp changes were visible while the system was functioning correctly. Hence preliminary indications are that model fit can be used as an advance failure alarm. However further analysis is required to estimate true and false positive rates, as well as the distribution of the obtained lead times. In particular, a larger annotated event base is required, where one can distinguish between failure of the system and failure of the monitoring platform.}

\section{Conclusion}\label{sec_concl}
We have presented a 2-layer model of system-level power consumption for {\Eurora}, a hybrid HPC installation containing CPUs, GPUs and MICs. The model takes as input workload parameters, namely job names and resources allocated to each job. It first computes  total power for computing components using Support Vector Regression and then predicts total power at system level, including also networking, IO and other components, by employing a linear model. The approach achieves very good performance on test data, with errors under 3\% for the month of October 2014.  \revised{The methodology can be easily applied to other systems since the data types used are generally available in most HPC systems.}

We have discussed several applications of our predictions. The first is optimizing or capping power usage for HPC systems, where prediction of system power can be employed to explore various scheduling and resource allocation schemes. Secondly, the model can be employed in predicting abnormal behaviors. We have indications that model fit can be used as an early signal of failure, however further analysis is required in this direction, with more complete annotated data. This analysis, together with power optimization, will be undertaken in future work.

\section*{Acknowledgment}

BigQuery analysis was carried out through a generous Cloud Credits grant from Google. The {\Eurora} database was developed by Prof. L. Benini and his group in the University of Bologna in collaboration with the HPC group at {\Cineca}. We are particularly grateful to Dr. A. Bartolini for useful discussions regarding the data and the system as a whole and to Dr. E. Rossi and Dr. C. Cavazzoni for providing access to the {\Cineca} systems. We acknowledge the {\Cineca} ISCRA  PACNA and PM-HPC  awards allowing access to HPC resources. \revised{This work has been partially funded by the European project SoBigData Research Infrastructure --- Big Data and Social Mining Ecosystem under the INFRAIA-H2020 program (grant agreement 654024).}

\bibliographystyle{IEEEtran}
\bibliography{refs}

\end{document}